\documentclass[journal]{IEEEtran}
\usepackage{amsmath,amsfonts}
\usepackage{array}
\usepackage{textcomp}
\usepackage{stfloats}
\usepackage{url}
\usepackage{verbatim}
\usepackage{graphicx}
\usepackage{cite}
\usepackage{outlines}

\usepackage{subcaption}
\usepackage{color}
\graphicspath{{./figures/}}
\usepackage{float}
\usepackage{amssymb} 
\usepackage{booktabs}
\usepackage[hidelinks]{hyperref}%

\begin{document}

\title{Reinforcement Learning-based Home Energy Management with Heterogeneous Batteries and Stochastic EV Behaviour}
\author{Meng Yuan,~\IEEEmembership{Member,~IEEE,} Ye Wang,~\IEEEmembership{Senior Member,~IEEE,} Xinghuo Yu,~\IEEEmembership{Fellow,~IEEE,} \\Torsten Wik,~\IEEEmembership{Senior Member,~IEEE,} Changfu Zou,~\IEEEmembership{Senior Member,~IEEE}
  \thanks{This work was supported  by European Union's Horizon Europe programme under the Marie Skłodowska-Curie Actions Postdoctoral Fellowships (Grant No.~101110832), the Australian Research Council via the 2022 Discovery Early Career Researcher Award (DECRA) (Grant No.~DE220100609), and the Science and Technology Innovation program of Hunan Province (Grant No.~2024RC4004).}%
  \thanks{Meng Yuan is with the School of Engineering and Computer Science, Victoria University of Wellington, Kelburn 6012, New Zealand, and was with the Chalmers University of Technology, 41296, Sweden. (E-mail: meng.yuan@ieee.org)}
  \thanks{Ye Wang is with the School of Mathematics and Statistics, The University of Melbourne, Parkville, VIC, 3010, Australia. (E-mail: ye.wang@unimelb.edu.au)}
  \thanks{Xinghuo Yu is with the School of Engineering, RMIT University, Melbourne, VIC, 3046, Australia. (E-mail: xinghuo.yu@rmit.edu.au)}
  \thanks{Torsten Wik and Changfu Zou are with the Department of Electrical Engineering, Chalmers University of Technology, Gothenburg, 41296, Sweden. (E-mails: torsten.wik@chalmers.se; changfu.zou@chalmers.se)}
}

\maketitle
\begin{abstract}

  The widespread adoption of photovoltaic (PV), electric vehicles (EVs), and stationary energy storage systems (ESS) in households increases system complexity while simultaneously offering new opportunities for energy regulation. However, effectively coordinating these resources under uncertainties remains challenging. This paper proposes a novel home energy management framework based on deep reinforcement learning (DRL) that can jointly minimise energy expenditure and battery degradation while guaranteeing occupant comfort and EV charging requirements. Distinct from existing studies, we explicitly account for the heterogeneous degradation characteristics of stationary and EV batteries in the optimisation, alongside stochastic user behaviour regarding arrival time, departure time, and driving distance. The energy scheduling problem is formulated as a constrained Markov decision process (CMDP) and solved using a Lagrangian soft actor-critic (SAC) algorithm. This approach enables the agent to learn optimal control policies that enforce physical constraints, including indoor temperature bounds and target EV state of charge upon departure, despite stochastic uncertainties. Numerical simulations over a one-year horizon demonstrate the effectiveness of the proposed framework in satisfying physical constraints while eliminating thermal oscillations and achieving significant economic benefits. Specifically, the method reduces the cumulative operating cost substantially compared to two standard rule-based baselines while simultaneously decreasing battery degradation costs by 8.44\%.

\end{abstract} %


\section{Introduction}
\IEEEPARstart{R}{esidential} buildings contribute a significant and growing share of global energy consumption, accounting for roughly 40\% of total energy use and a substantial portion of electricity demand \cite{farahmandfar2025towards}. This rising energy demand, coupled with a broader electrification, such as the wide use of electric vehicles (EVs), is driving interest in smarter home energy management systems (HEMS). Meanwhile, the increase of rooftop photovoltaic (PV) systems, home battery energy storage systems (ESS), and EVs offers new opportunities for households to meet their energy needs cheaper and more sustainably. Combining flexible EV charging and discharging with local resources, like solar PV and battery storage, transforms homes into active energy hubs for the household or microgrid \cite{sangswang2020optimal}. However, coordinating multiple devices like batteries and EVs under various operational constraints can be challenging. The system must schedule these devices to reduce costs while respecting user comfort, device health, and the unpredictability of EV usage, and these complexities motivate the need for advanced control strategies \cite{nguyen2013joint, gholinejad2022hierarchical, islam2021coordinating, saber2022transactive}.

Early HEMS research largely employed rule-based or heuristic controllers, with implementations targeting time-of-use bill reduction through demand shifting and peak shaving \cite{hussain2020heuristic}. Later, optimisation-based methods such as mixed-integer programming co-optimised PV, battery operation, and load schedules to minimise electricity cost \cite{thomas2018optimal}. In \cite{lu2020optimal}, a multi-objective HEMS that jointly minimises electricity expenditure and deviations from thermal-comfort set-points was formulated, which permits limited temperature excursions when economically justified. Moreover, existing works often overlook battery dynamics, relying instead on simple operational constraints to indirectly limit cycling ageing. While research, such as \cite{farzin2016practical}, has improved scheduling realism by including battery wear costs, existing literature often relies on simplified or generalised degradation penalty terms. These approaches typically overlook the intricate electrochemical differences between varying battery technologies, aiming instead for a broad approximation of ageing costs suitable for single-battery systems.

This limitation becomes critical with the growing maturity of vehicle-to-grid (V2G) and vehicle-to-home (V2H) technologies. These technologies allow EVs to serve as high-capacity, dispatchable energy resources for the HEMS \cite{al2023optimal}. By discharging during high-cost periods, storing excess solar energy, or providing backup power during an outage, the EV can significantly reduce electricity bills and increase energy resilience \cite{erdinc2014smart, wu2016stochastic}. However, the degradation characteristics of an EV battery are fundamentally different from those of a stationary battery. A HEMS that ignores this heterogeneity may impose frequent and aggressive charge-discharge cycles on the EV battery, leading to accelerated wear and potentially limiting the life of the vehicle \cite{huang2022investigation}.

Recently, reinforcement learning (RL) has emerged as a prominent methodology for HEMS, primarily due to its capacity to learn near-optimal control policies directly from operational data \cite{yu2019deep, mansour2025deep, lissa2021deep}. In \cite{yu2019deep}, deep RL is applied to solve a smart home energy management problem, resulting in notable energy cost savings while maintaining indoor thermal control. In \cite{zenginis2022smart}, a multi-agent DDPG framework using data clustering was presented to train specialised agents that each handle specific operational scenarios, leading to more efficient energy cost. However, existing RL-based HEMS methods frequently overlook real-world complexities by modelling EV charging behaviour with simplistic or deterministic assumptions, such as fixed arrival and departure schedules, thereby failing to capture the stochastic nature of user behaviour \cite{wang2016predictive, niu2024uncertainty}.

In response to these challenges, this paper proposes a novel HEMS framework based on deep reinforcement learning (DRL). Standard reinforcement learning formulations typically handle system constraints by adding static penalty terms to the reward function. This approach, however, often necessitates tedious manual tuning and leads to a sensitive trade-off between cost minimisation and constraint satisfaction. We therefore formulate the energy scheduling problem as a constrained Markov decision process (CMDP). This framework explicitly decouples the economic objective from operational limits, enabling the co-optimisation of ESS usage, EV charging schedules, and heating, ventilation, and air conditioning (HVAC) and appliance operation. Specifically, the primary objective is to minimise the net grid electricity and battery degradation costs while strictly satisfying occupant comfort constraints and EV charging requirements.

To solve this problem, we employ a Lagrangian-based soft actor-critic (SAC) algorithm. This method is chosen for its ability to handle continuous control variables and adapt to stochastic dynamics. By introducing a dual-variable mechanism, this method allows the agent to automatically adapt penalty multipliers during training, thereby learning a control policy that minimises operational costs while intrinsically respecting comfort limits. Furthermore, to bridge the gap between theoretical simulation and real-world application, we validate the proposed framework in a high-fidelity learning environment tailored to a Swedish household scenario. This environment is designed to capture realistic complexities, including heterogeneous battery degradation dynamics and stochastic EV user behaviours derived from local travel survey data.

To the best of the author's knowledge, this is the first HEMS study to jointly account for different battery degradation dynamics and uncertain EV behaviour within a Lagrangian SAC framework. The main contributions of this work are summarised as follows:

\begin{enumerate}
  \item To accurately capture the degradation costs associated with ageing and cycling, a semi-empirical capacity fade model distinguishing between lithium iron phosphate (LFP) for stationary storage and nickel manganese cobalt (NMC) for EVs is established and incorporated into the optimisation objective. This differs from existing works that typically ignore degradation or assume uniform characteristics for all storage units.

  \item A realistic stochastic modelling approach for uncertain EV user behaviour is integrated into the learning environment. By sampling arrival times, departure times, and daily driving distances from distributions fitted to Swedish national travel survey data, the learned policy demonstrates superior robustness against unpredictable vehicle availability compared to deterministic baselines.

  \item A constrained SAC algorithm based on the Lagrangian relaxation method is developed to solve the multi-objective optimisation problem. By automatically adapting dual variables, the proposed framework enforces occupant thermal comfort and EV charging requirement without the need for experience-driven weight tuning.
\end{enumerate}

\section{System Modelling}

\begin{figure}[tb]
  \centering
  \includegraphics[width=\linewidth]{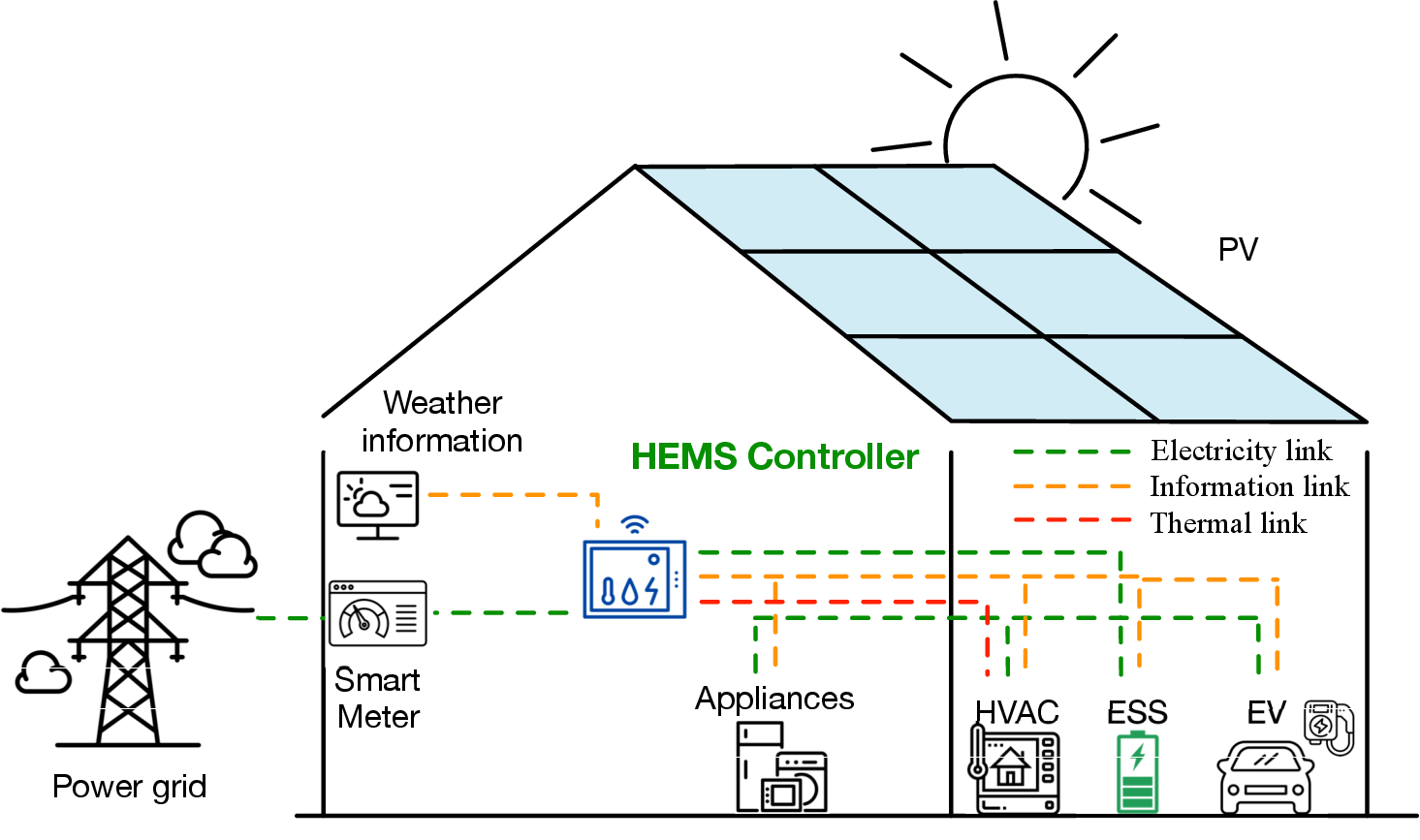}
  \caption{Structure of the investigated home with a smart energy management system.}
  \label{fig:schematic_system}
\end{figure}

This section presents the detailed modelling of the smart home energy management system, the structure of which is illustrated in Fig.~\ref{fig:schematic_system}. We first present detailed models for household appliances, categorised based on their operational flexibility. Subsequently, we describe models for local energy resources, including the PV generation subsystem, the HVAC system, the ESS, and the EV. Finally, a power balancing constraint is formulated to ensure the balance between total energy supply and demand within the home at every time step. These models collectively form the basis for smart home energy scheduling and optimisation.

\subsection{Model of Ordinary Residential Appliances}

In this work, ordinary residential loads are classified into three categories based on their operational flexibility and scheduling constraints. The first category comprises non-shiftable loads. These appliances operate according to strict, user-defined time windows and provide no flexibility for temporal rescheduling by the HEMS. The second category consists of time-shiftable but uninterruptible appliances (SUA), which offer temporal flexibility by allowing their start time to shift within a user-defined window. However, once an SUA initiates its operational cycle, it cannot be interrupted until the task is complete, as observed in dishwashers, washing machines, and electric ovens. The final category includes time-shiftable and interruptible appliances (SIA), which provide the highest level of flexibility. Their operation can be postponed, paused, or resumed, thereby allowing for precise control in response to grid conditions or electricity prices.

Let $L_i^{\text{on}}$ denote the required operation duration for appliance $i$. We define the binary operation status $z_i(t) \in \{0,1\}$ and the start-up indicator $\mu_{i}(t) \in \{0,1\}$. We assume all devices are initially OFF, i.e., $z_{i}(t_{i}^b -1) = 0$. For both SUA and SIA categories, the start-up logic and minimum operation duration are governed by identical physical constraints. The general formulation for any shiftable appliance $i$ is given as:
\begin{subequations} \label{eq:general_logic}
  \begin{align}
    & \mu_{i}(t) \geq z_{i}(t) - z_{i}(t-1), \label{eq:gen_mu_a}\\
    & \mu_{i}(t) \leq z_{i}(t), \label{eq:gen_mu_b}\\
    & \mu_i(t)\leq 1- z_{i}(t-1),  \label{eq:gen_mu_c}\\
    & z_{i}(t+k) \geq \mu_{i}(t),\; \forall k \in [1, L_{i}^{\text{on}}-1], \label{eq:gen_min_on}
  \end{align}
\end{subequations}
where constraints \eqref{eq:gen_mu_a}--\eqref{eq:gen_mu_c} define $\mu_{i}(t) = 1$ if and only if a start-up event occurs at time $t$. Constraint \eqref{eq:gen_min_on} enforces the minimum on-time $L_{i}^{\text{on}}$.

For the SUA category, the task must be executed exactly once within the admissible window $\mathcal{T}_i \triangleq \{t_i^{b},\ldots,t_i^{e}-L_i^{\mathrm{on}}\}$. Thus, in addition to \eqref{eq:general_logic}, the SUA is subject to:
\begin{equation} \label{eq:SUA_constraint}
  \sum_{t\in \mathcal{T}_i} \mu_{i}(t) = 1.
\end{equation}

For the SIA category, the appliance operates across multiple disjoint periods $j = 1, \ldots, M_{i}$. Let $\mathcal{T}_{i,j}^{\text{on}} \triangleq \{t_{i,j}^{b},\ldots,t_{i,j}^{e}-L_i^{\mathrm{on}}\}$ and $\mathcal{T}_{i,j}^{\text{off}} \triangleq \{t_{i,j}^{b},\ldots,t_{i,j}^{e}-L_i^{\mathrm{off}}\}$ denote the admissible start and shut-down times for period $j$, respectively. The SIA must start exactly once in each period $j$, satisfying:
\begin{equation} \label{eq:SIA_start}
  \sum_{t\in \mathcal{T}_{i,j}^{\text{on}}} \mu_{i}(t) = 1, \; \forall j \in [1,M_i].
\end{equation}

Additionally, to ensure the minimum downtime $L_{i}^{\text{off}}$, we define the shut-down indicator $\nu_{i}(t)$ and impose the following constraints for all $t \in \mathcal{T}_{i,j}^{\text{off}}$:
\begin{subequations} \label{eq:SIA_off}
  \begin{align}
    & \nu_{i}(t) \geq z_{i}(t-1) - z_{i}(t),\\
    & \nu_{i}(t) \leq z_{i}(t-1),\\
    & \nu_i(t)\leq 1- z_{i}(t), \\
    & \sum_{t \in \mathcal{T}_{i,j}^{\text{off}}} \nu_{i}(t) = 1,\; \forall j \in [1,M_i], \label{eq:SIA_stop_once}\\
    & 1-z_{i}(t+k)\geq \nu_{i}(t), \; \forall k \in [0, L_{i}^{\text{off}}-1], \label{eq:SIA_min_off}
  \end{align}
\end{subequations}
where constraint \eqref{eq:SIA_stop_once} ensures the appliance shuts down exactly once per period, and \eqref{eq:SIA_min_off} guarantees the minimum off-time after shut-down.

The state transition logic must satisfy the boundary conditions within the operating range $t\in [t_i^b, t_i^e]$:
\begin{subequations}
  \begin{align}
    & z_{i}(t) - z_{i}(t-1) = \mu_{i}(t)  -\nu_{i}(t),  \\
    & \sum_{t=t_i^b}^{t=t_i^e} \left( \mu_i(t) - \nu_i(t) \right) = z_{i}(t_{i}^{e}) - z_{i}(t_{i}^{b}), 
  \end{align}
\end{subequations}

Finally, ordinary appliances are subject to the power constraint:
\begin{align}
  z_{i} (t) P_i^{\min} &\leq P_{i}(t) \leq z_{i}(t) P_i^{\max}.\label{eq:power_ORA}
\end{align}

\subsection{PV Generation Subsystem}

Let $\bar{P}^{\text{PV}}(t)$ denote the available PV power at time $t$, and let $P^{\text{PV}}_{\max}$ be the rated capacity limit of the inverter. The dispatched PV output ${P}^\text{PV}(t)$ satisfies
\begin{equation}
  0 \leq {P}^\text{PV}(t) \leq \min\{\bar{P}^{\text{PV}}(t), P^{\text{PV}}_{\max} \},
\end{equation}
where $P^{\text{PV}}_{\max}$ is the maximum available PV power estimated from historical data. The generated power, $P^{\text{PV}}(t)$, is managed by the EMS to supply the household load or to be stored in the ESS or the EV battery.

\subsection{HVAC System}

In a home environment, the HVAC system is a major energy-consuming appliance and is responsible for maintaining the indoor temperature within a prescribed comfort range. It typically provides both heating and cooling functionalities. A commonly adopted discrete-time thermal model for residential buildings is given by \cite{yu2019deep,lu2020optimal}:
\begin{equation}
  T_{\text{in}}(t+1) = \varepsilon T_{\text{in}}(t) + \left(T_{\text{out}}(t) + \delta(t) \alpha_{\text{T}} P^{\text{HVAC}}(t)\right)(1-\varepsilon),
\end{equation}
where $T_{\text{in}}$ and $T_{\text{out}}$ denote the indoor and outdoor temperatures, respectively, and $\varepsilon \in (0,1)$ characterises the thermal inertia (or heat retention) of the building envelope. The term $P^{\text{HVAC}}(t)$ represents the electrical power consumed by the HVAC system and is constrained by its rated power capacity $P^{\text{HVAC}}_{\max}$:
\begin{equation}
  0 \leq P^{\text{HVAC}}(t) \leq P^{\text{HVAC}}_{\max}.
\end{equation}
The parameter $\alpha_{\text{T}}$ is the thermal conversion efficiency coefficient that maps electrical power into indoor temperature change. The operating mode of the HVAC system is determined by $\delta(t)$, defined as
\begin{equation}
  \delta(t) =
  \begin{cases} +1 & \text{heating mode}, \\ -1 & \text{cooling mode}, \\ 0 & \text{HVAC off}.
  \end{cases}
\end{equation}

\subsection{Energy Storage System}

In residential settings, PV systems convert sunlight into electricity, but their power output is inherently unstable due to the intermittent nature of solar irradiance and stochastic weather conditions. An ESS can buffer these fluctuations, improve PV self-consumption, and arbitrage time-of-use tariffs.

Let $\text{SoC}^{\text{ESS}}(t)$ be the state of charge (SoC) of the stationary battery at time $t$, and let $E^{\text{ESS}}$ denote its energy capacity.

The SoC dynamics of ESS can be formulated as \cite{yu2019deep}:
\begin{equation}
  \text{SoC}^\text{ESS}(t+1) = \text{SoC}^\text{ESS}(t) + \frac{\eta^\text{ESS,c}P^\text{ESS,c}(t) \Delta t}{E^{\text{ESS}}} + \frac{ P^\text{ESS,d}(t) \Delta t}{\eta^\text{ESS,d}E^{\text{ESS}}},
\end{equation}
where $\eta^{\text{ESS,c}}$ and $\eta^{\text{ESS,d}}$ are charging and discharging efficiencies, respectively. We adopt the sign convention $P^{\text{ESS,c}}(t) \geq 0$ for charging power and $P^{\text{ESS,d}}(t) \leq 0$ for discharging power. Time is discretised with a sampling interval of $\Delta t=1\,\mathrm{h}$.

The SoC of ESS is constrained by
\begin{equation}
  \text{SoC}^\text{ESS}_\text{min}\leq\text{SoC}^\text{ESS}(t)\leq\text{SoC}^\text{ESS}_\text{max}.
\end{equation}

To prevent simultaneous charging and discharging within a time step and ensure the power remains within bounds, we enforce the following constraint with a binary variable $z^{\text{ESS}}(t)\in \{0,1\}$:
\begin{equation}
  0 \leq P^\text{ESS,c}(t) \leq  z^{\text{ESS}}(t) P^\text{ESS,c}_{\max},
\end{equation}
\begin{equation}
  -\big( 1-z^{\text{ESS}}(t) \big) P^\text{ESS,d}_{\max}  \leq P^\text{ESS,d}(t) \leq  0.
\end{equation}

\subsection{EV Model}

We consider an EV parked at home from $t_{\text{arr}}$ to $t_{\text{dep}}$. Let $\text{SoC}^{\text{EV}}(t)$ denote the SoC at time $t$ and $E^{\text{EV}}$ the battery capacity. The SoC dynamics during the parked period $t \in [t_{\text{arr}}, t_{\text{dep}}]$ are governed by
\begin{equation}
  \mathrm{SoC}^{\mathrm{EV}}(t{+}1)
  = \mathrm{SoC}^{\mathrm{EV}}(t)
  + \frac{\eta^{\mathrm{EV,c}} P^{\mathrm{EV,c}}(t) \Delta t}{E^{\mathrm{EV}}}
  + \frac{P^{\mathrm{EV,d}}(t) \Delta t}{\eta^{\mathrm{EV,d}}\,E^{\mathrm{EV}}},
\end{equation}
where $\eta^{\mathrm{EV,c}}$ and $\eta^{\mathrm{EV,d}}$ are charging and discharging efficiencies, respectively. We adopt the convention that $P^{\mathrm{EV,c}}(t) \ge 0$ for charging and $P^{\mathrm{EV,d}}(t) \le 0$ for discharging.

The SoC is required to stay within the following range:
\begin{equation}
  \mathrm{SoC}^{\mathrm{EV}}_{\min}\le \mathrm{SoC}^{\mathrm{EV}}(t)\le \mathrm{SoC}^{\mathrm{EV}}_{\max}.
\end{equation}

Similar to the ESS, we require that the battery cannot charge and discharge simultaneously while respecting power limits:
\begin{subequations}
  \begin{align}
    &0 \leq P^{\mathrm{EV,c}}(t) \leq z^{\mathrm{EV}}(t)\,P^{\mathrm{EV,c}}_{\max},\\
    &-\,\big(1{-}z^{\mathrm{EV}}(t)\big) P^{\mathrm{EV,d}}_{\max} \le P^{\mathrm{EV,d}}(t) \le 0,\\
    &z^{\mathrm{EV}}(t)\in\{0,1\}.
  \end{align}
\end{subequations}

When the EV is not at home, the power obeys
\begin{equation}
  P^{\mathrm{EV,c}}(t)=P^{\mathrm{EV,d}}(t)=0,\quad \forall t\notin [t_{\text{arr}}, t_{\text{dep}}].
\end{equation}

The arrival and departure conditions are given by
\begin{equation}
  \mathrm{SoC}^{\mathrm{EV}}(t_{\mathrm{arr}})=\mathrm{SoC}^{\mathrm{EV}}_{\mathrm{arr}},\quad
  \mathrm{SoC}^{\mathrm{EV}}(t_{\mathrm{dep}})\ge \mathrm{SoC}^{\mathrm{EV}}_{\mathrm{req}}.
\end{equation}

\subsection{Power Balancing}

The power supply must be balanced within the home environment, ensuring that the total supply meets the demand. This can be realised by enforcing the following equality constraint:
\begin{align}
  P^{\text{grid}}(t)+P^{\text{PV}}(t) - P^{\text{ESS,d}}(t) - P^{\text{EV,d}}(t) = \sum_{i=1}^{N}z_{i}(t)P_{i} \nonumber \\
  + P^{\text{ESS,c}}(t) +P^{\text{EV,c}}(t) +  P^{\text{HVAC}}(t), \label{eq:power_balance}
\end{align}
where $N$ is the total number of ordinary residential appliances, and $P^{\text{grid}}$ represents the net power exchanged with the grid, which is defined as positive for import and negative for export.

\section{Statistical Modelling of Uncertainty in Home Energy Management}

Although electricity prices and PV generation are often treated as stochastic, day-ahead prices in Nordic markets like Sweden are published in advance, and PV forecasting is a well-established research area. Consequently, this work treats them as deterministic to focus on the significant stochasticity of EV user behaviour. Specifically, we model three key uncertainties associated with the EV: (i) arrival time at home, (ii) departure time, and (iii) daily driving distance, which determines the initial SOC upon arrival home.

\subsection{Travel Behaviour Analysis}

To capture the uncertainties related to EV user behaviour, we analyse the departure and arrival times based on data from a travel survey report by Region Stockholm \cite{regionstockholm2020}.

Departure times from home are typically right-skewed. To capture the characteristic steep initial rise and long right tail of this distribution, we employ the log-normal distribution defined as
\begin{equation}\label{eq:time_departure}
  f\!\left(t_{\text{dep}} \mid \mu_{\text{d}}, \sigma_{\text{d}}, \alpha_{\text{d}}\right)
  = \frac{1}{\sqrt{2\pi}(t_{\text{dep}}-\mu_{\text{d}})\,\sigma_{\text{d}}}
  e^{
    -\frac{\left(\log\!\frac{t_{\text{dep}}-\mu_{\text{d}}}{\alpha_{\text{d}}}\right)^{2}}{2\sigma_{\text{d}}^{2}}
  },
\end{equation}
where $\mu_{\text{d}}$, $\sigma_{\text{d}}$, $\alpha_{\text{d}}$ are the location, scale, and shape parameters, respectively.

\begin{figure}[tb]
  \centering
  \includegraphics[width=0.9\linewidth]{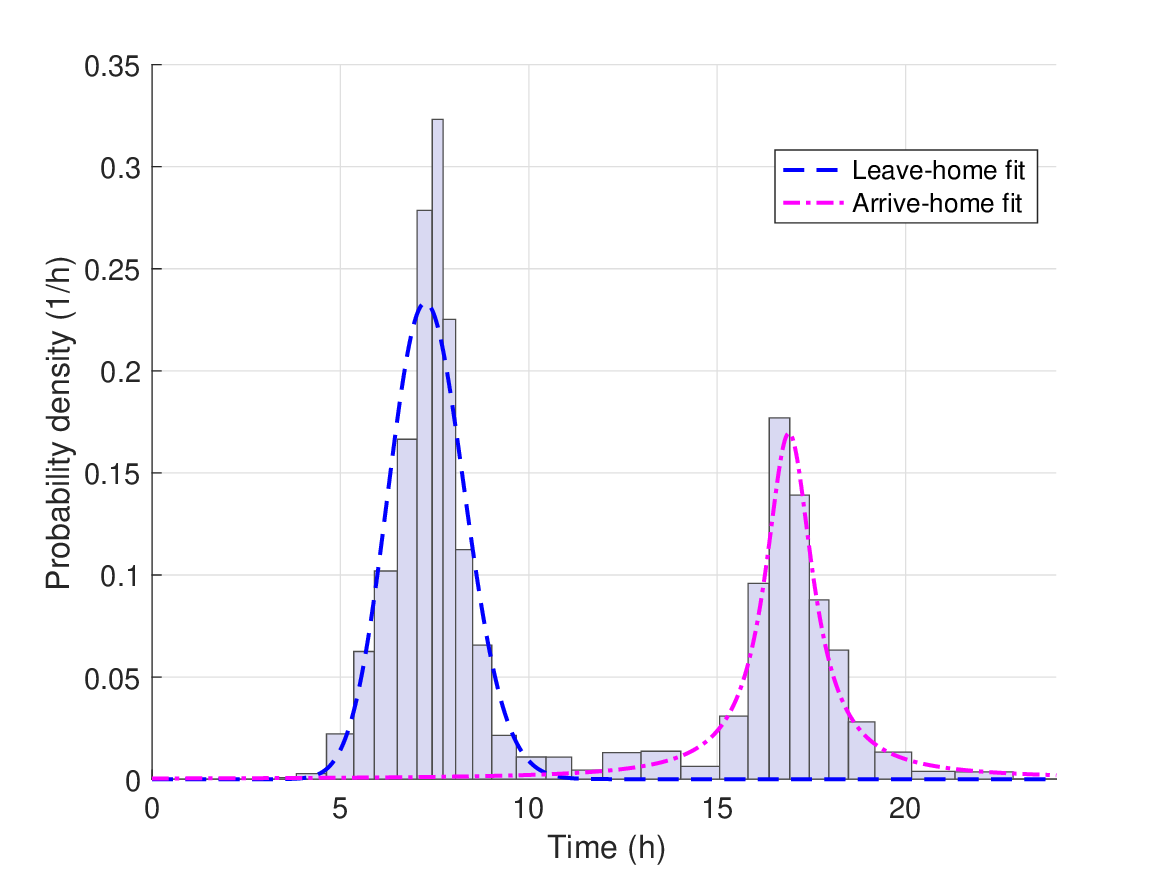}
  \caption{The distributions of leaving and arrival home time.}
  \label{fig:uncertainty_time}
\end{figure}

Arrival times at home tend to peak sharply in the evening and exhibit heavy tails. This behaviour is modelled by the Cauchy distribution:
\begin{equation} \label{eq:time_arrival}
  f\!\left(t_{\text{arr}} \mid s_{\text{a}}, \gamma_{\text{a}}\right)
  = \frac{1}{\pi} \,
  \frac{\gamma_{\text{a}}}{\left(t_{\text{arr}} - s_{\text{a}}\right)^{2} + \gamma_{\text{a}}^{2}},
\end{equation}
where $s_{\text{a}}$ and $\gamma_{\text{a}}$ are the location and scale parameters that determine the central tendency and spread of the distribution. The specific parameters for \eqref{eq:time_departure} and \eqref{eq:time_arrival} are summarised in Table~\ref{tab:time_parameters} in Appendix. The resulting fitted distribution is visualised in Fig.~\ref{fig:uncertainty_time}.

\subsection{Battery SoC Uncertainty Analysis}

The state of charge (SoC) of the EV upon arrival is mainly influenced by driving distance, traffic conditions, and driving behaviour \cite{zhu2024predicting}. In this work, we treat the arrival SoC as a random variable determined by the daily travel distance
\begin{equation}
  \text{SoC}^{\text{EV}}(t_{\text{arr}}) = \text{SoC}^{\text{EV}}(t_{\text{dep}}) - \frac{\eta^{\text{EV}}L^{\text{EV}}}{E^{\text{EV}}},
\end{equation}
where $E^{\text{EV}}$ denotes the battery capacity of the EV in kWh, and $\eta^{\text{EV}}$ represents the average energy consumption in kWh/km. The daily travel distance $L^{\text{EV}}$ is modelled based on data from a report by Jönköping municipality in Sweden \cite{jonkopings_kommun_2024}. The probability distribution function (PDF) of the travel distance is approximated using a Gaussian mixture model given by
\begin{equation}
  f(L^{\text{EV}}) = \sum_{i=1}^{n}  \frac{A_{i}}{\sqrt{2\pi}\sigma_i} e^{-\frac{(L^{\text{EV}}-\mu_i)^2}{2\sigma_i^2}}.
\end{equation}

The resulting PDF with $n=3$ components is illustrated in Fig.~\ref{fig:uncertainty_distance}, and the detailed fitting parameters are summarised in Appendix.

\begin{figure}[tb]
  \centering
  \includegraphics[width=0.9\linewidth]{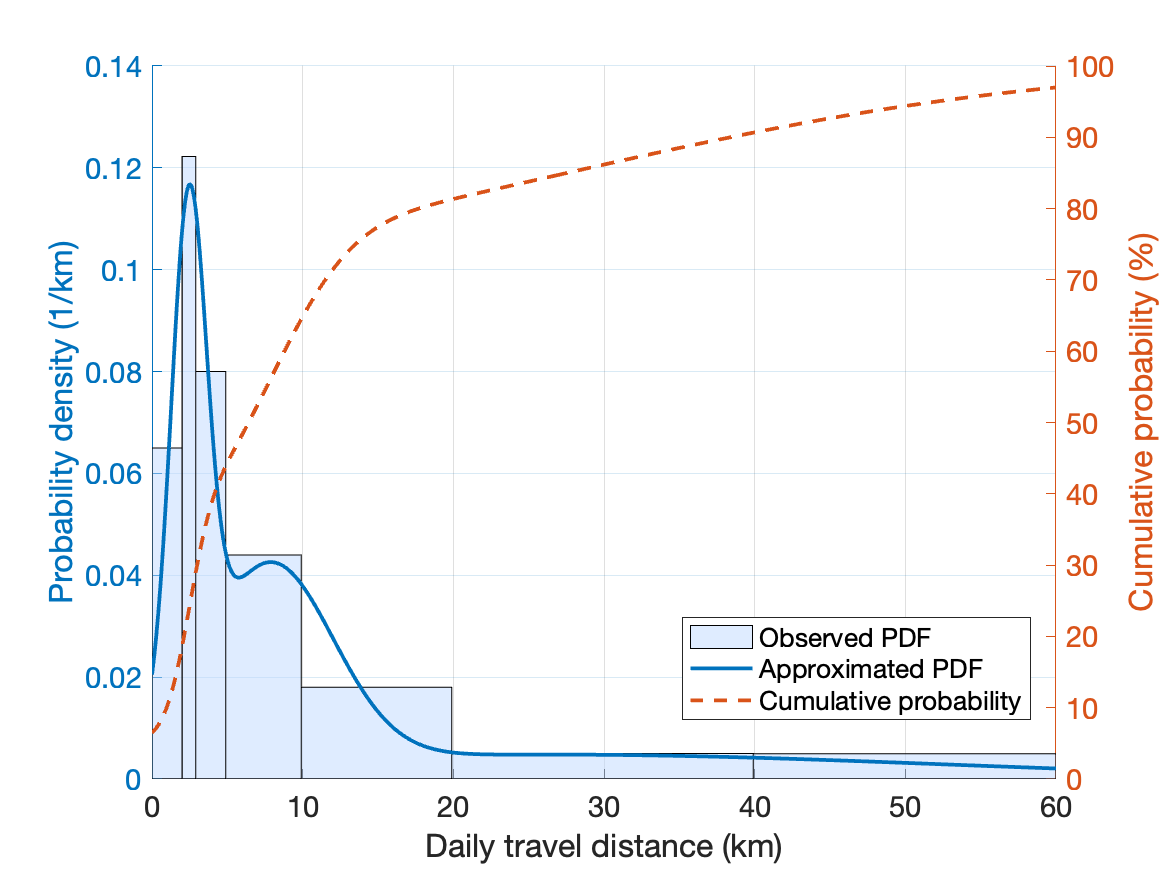}
  \caption{Daily travel distance distribution.}
  \label{fig:uncertainty_distance}
\end{figure}

\section{Problem Formulation and Cost Consideration}

In this work, we aim to minimise the expected total cost over the horizon, comprising the net electricity cost, a thermal-discomfort penalty, and the battery degradation cost. Accordingly, the stochastic multi-objective optimisation problem is formulated as follows:
\begin{align}\label{eq:problem_optimisation}
  & \min_{\varpi(t)} J(t) = \nonumber \sum_{t=1}^{H-1} \mathbb{E}_{\xi}\left\{ C_{\text{grid}}(t) +  C_{\text{comfort}}(t) +  C_{\text{deg}}(t) \right\}, \\
  & \text{s.t.} \; \eqref{eq:general_logic}-\eqref{eq:power_balance},
\end{align}
where $\varpi$ denotes the control policy mapping system states to actions, $\xi = \{ t_{\text{dep}}, t_{\text{arr}}, L^{\text{EV}} \}$ is the vector of stochastic parameters and $C_{\text{grid}}$ represents the cost related to the electricity tariff, expressed explicitly as
\begin{equation}
  C_{\text{grid}}(t)
  = \frac{r_b(t)-r_s(t)}{2}\,\lvert P^{\text{grid}}(t) \rvert
  + \frac{r_b(t)+r_s(t)}{2}\,P^{\text{grid}}(t),
\end{equation}
where $r_b$ and $r_s$ denote the buying and selling prices of electricity, respectively. While the grid cost is derived directly from market prices, the terms $C_{\text{comfort}}$ and $C_{\text{deg}}$ involve physical dynamics and user preferences, thus, their detailed models are presented in the following subsections.

\subsection{User Satisfaction Cost}

The user satisfaction cost accounts for both thermal comfort and the fulfilment of EV charging requirements. Let $T^{\text{ref}}_{\min}$ and $T^{\text{ref}}_{\max}$ denote the lower and upper bounds of the preferred indoor temperature range, respectively. To allow some slack, we introduce two nonnegative slack variables and have following constraints: 
\begin{align}
  T_{\mathrm{in}}(t) &\le T_{\max}^{\mathrm{ref}}(t) + v^{+}(t), \, v^+(t) \geq 0\\
  T_{\mathrm{in}}(t) &\ge T_{\min}^{\mathrm{ref}}(t) - v^{-}(t), \, v^-(t) \geq 0. \label{eq:slacks_nonneg}
\end{align}

Also, the SoC of EV at departure affects the user satisfaction, we model this by
\begin{equation}
  v_{\mathrm{dep}}(t) =
  \begin{cases}
    0, & t \neq t_{\mathrm{dep}}, \\[4pt]
    \max\!\big(0,\,
      \text{SoC}_{\mathrm{req}}^{\mathrm{EV}} - \text{SoC}^{\mathrm{EV}}(t)
    \big), & t = t_{\mathrm{dep}}.
  \end{cases}
  \label{eq:vdep}
\end{equation}

With this, the total discomfort cost is defined as:
\begin{equation}
  C_{\text{comfort}}(t) =
  w_c^+ v^+(t) + w_c^- v^-(t)
  +  w_d v_{\mathrm{dep}}(t), 
\end{equation}
where the weights $w_c^{+}$ and $w_c^{-}$ penalise overheating and under-heating, respectively, while $w_d$ penalises the deviation of the EV SoC from the user's expectation. %

\subsection{Battery Degradation Cost}

This work considers a combined energy system incorporating both stationary and EV batteries. The LFP battery is suitable for stationary storage due to its thermal stability, longevity, and lower cost. In contrast, ternary lithium batteries, primarily NMC, are widely used in EVs because their high energy density supports vehicle acceleration and range, despite having shorter lifespans and higher costs. These fundamental differences result in distinct degradation behaviours that significantly impact the energy management strategy. To address this, we explicitly incorporate heterogeneous degradation models for both battery types into the optimisation framework.

Battery ageing is primarily categorised into calendar ageing and cycle ageing. The capacity loss due to calendar ageing can be described by \cite{xu2023electric}
\begin{equation} \label{eq:loss_cal}
  \begin{split}
    Q_{\text{loss,cal}}(T,\text{SoC},t)  =\: & k_{\text{cal}}  \cdot \exp\left[     -\frac{E_a}{R T} \left( \frac{1}{T} - \frac{1}{T_{\text{ref}}} \right) \right.\\
      &  \left. + \frac{\alpha_{\text{cal}} F}{R} \left( \frac{U_a}{T} - \frac{U_{a,\text{ref}}}{T_{\text{ref}}} \right)
    \right] \sqrt{t}, 
  \end{split}
\end{equation}
where $k_{\text{cal}}$, $E_{a}$, and $\alpha_{\text{cal}}$ are fitting parameters, $R$ is the universal gas constant, $T$ is the battery temperature, and $F$ is the Faraday constant. The anode-to-reference potential $U_{a}$ is a function of SoC given by
\begin{align}
  U_a(\text{SoC}) =\ & 0.6379 + 0.5416 \cdot \exp \left( -305.5309 \cdot x_a(\text{SoC}) \right) \nonumber\\
  & + \sum_{i=1}^{4} a_i \cdot \tanh\left( \frac{x_a(\text{SoC}) - b_i}{c_i} \right),
\end{align}
where \( a_i, b_i, c_i \) are empirical fitting parameters summarised in Table~\ref{tab:ua_params} according to \cite{safari2011modeling}. The variable $x_{a}$ represents the lithiation fraction of graphite, modelled linearly as:
\begin{equation}
  x_a(\text{SoC}) = x_{a,0} + \text{SoC} \cdot (x_{a,100} - x_{a,0}),
\end{equation}
where \( x_{a,0} = 0.0085 \) and \( x_{a,100} = 0.78 \) represent the lithiation fractions at 0\% and 100\% SoC, respectively \cite{schimpe2018comprehensive}. The reference potential $U_{a,\text{ref}}$ is calculated at 50\% SoC and $T_{\text{ref}} = 25^\circ\mathrm{C}$.

For cycle-induced degradation, we employ the following semi-empirical model \cite{xu2023electric}:
\begin{multline}
  \label{eq:loss_cyc}
  Q_{\text{loss,cyc}} = k_{\text{cyc}} (\alpha_{c1} \text{DOD} + \alpha_{c2}) (\alpha_{c3} C_{\text{rate}} + \alpha_{c4}) \\
  \cdot (\alpha_{c5}  (T - T_{\text{ref}})^2 + \alpha_{c6})  N_{\text{EFC}},
\end{multline}
where $N_{\text{EFC}}$ is the equivalent full-cycle number, and the terms involving $\alpha_{ci}$ capture the effects of depth of discharge (DOD), C-rate, and temperature.

To compute the degradation cost, the incremental capacity loss for battery $k \in \{\text{ESS}, \text{EV}\}$ over one control step $[t,t+\Delta t]$ is calculated as:
\begin{align}
  \Delta Q_{\text{loss}}^{(k)}(t) & = \Delta Q_{\text{loss, cal}}^{(k)}(t) + \Delta Q_{\text{loss, cyc}}^{(k)}(t) \nonumber\\
  & = Q_{\text{loss, cal}}^{(k)}(T, \text{SoC}, t+\Delta t) - Q_{\text{loss, cal}}^{(k)}(T, \text{SoC}, t) \nonumber \\
  & \quad + k_{\text{cyc}}^{(k)} f_{\text{DOD}}^{(k)} f_{\text{C-rate}}^{(k)} f_{T}^{(k)} \Delta N_{\text{EFC}}^{(k)}(t).
\end{align}

Here, $\Delta N_{\text{EFC}}^{(k)}(t)$ denotes the incremental equivalent full cycles, computed from energy throughput:
\begin{equation}
  \Delta N_{\text{EFC}}^{(k)}(t) = \frac{|P^{(k), c/d}(t)| \Delta t}{2 E^{(k)} }, 
\end{equation}
where $E^{(k)}$ is the nominal energy of the batteries. We use the effective stepwise DOD for computing cycle degradation:
\begin{equation}
  \text{DOD}(t) = \min \{1, \Delta N_{\text{EFC}}^{(k)}(t) \}.
\end{equation}

While methods such as rain-flow counting offer higher precision for irregular cycles, this stepwise approximation allows for efficient integration into the optimisation framework. Finally, the degradation cost is given by:
\begin{equation}
  C_{\text{deg}}(t) = \sum_{k} w_{\text{deg}}^{(k)} \Delta Q_{\text{loss}}^{(k)}(t), 
\end{equation}
where $w_{\text{deg}}^{(k)}$ weights the economic impact of capacity loss, with $w_{\text{deg}}^{\text{EV}} > w_{\text{deg}}^{\text{ESS}}$ reflecting the higher replacement cost of the EV battery.

\section{RL-based Optimisation Formulation}\label{sec:RL_algorithm}

The optimisation problem formulated in Section IV involves complex nonlinear degradation dynamics and stochastic EV behaviour. To address this, we employ a deep reinforcement learning approach, which enables the agent to learn optimal control policies directly from the environment, effectively managing continuous variables and uncertainties without the need for explicit forecasting.

\subsection{MDP Formulation}

The hourly energy scheduling problem is formulated as a CMDP, as the system dynamics determining the next-hour state are fully specified by the current state and the applied control. Specifically, the indoor temperature at $t+1$ depends on the current indoor temperature, outdoor temperature, and HVAC actuation. The SoC of the ESS and EV evolve based on their current levels and charge or discharge power, while appliance states update according to their operational constraints. Uncertainties such as EV user behaviour are modelled as stochastic disturbances that are conditionally independent of the past, given the current time context. We augment the state with time-of-day encoding and availability flags to ensure the transition $P(s_{t+1}|s_t,a_t)$ is approximately Markovian. %

The objective is to minimise the expected cumulative operating cost over the horizon $H$, subject to comfort and device limits. This yields a discounted CMDP defined by the tuple $\mathcal{M}=(\mathcal{S},\mathcal{A},\mathcal{P},\mathcal{R},\gamma,g,d)$, where $\mathcal{S}$ is the set of environment states and $\mathcal{A}$ is the set of actions. The transition probability function is denoted by $\mathcal{P}(s'| s,a) \in [0,1]$. The reward function $R$ aggregates economic and degradation terms, while $g$ represents the constraint cost functions with budget $d$.

\subsubsection{Environment State} The electricity selling price is assumed to be proportional to the buying price with a constant factor $\sigma \le 1$. Thus, $r_s(t)$ is not included as an independent state variable. The environment state at time $t$ comprises the time index $t$, indoor temperature $T_{\text{in}}(t)$, outdoor temperature $T_{\text{out}}(t)$, the SoC levels of the stationary battery and the EV, the available PV power $P^{\text{PV}}(t)$, the electricity buying price $r_b(t)$, and the EV presence indicator $z^{\text{EV}}(t)$. The environment state at time $t$ is thus described as $s_{t}  = \{T_{\text{in}}(t), T_{\text{out}}(t), \text{SoC}^{\text{ESS}}(t), \text{SoC}^{\text{EV}}(t), P^{\text{PV}}(t), r_b(t), z^{\text{EV}}(t)\}.$

\subsubsection{Action} The action space of the proposed agent consists of continuous variables for power regulation and discrete variables for appliance scheduling. These include the HVAC power, ESS charging/discharging powers, EV charging/discharging powers, and the scheduling decisions for each appliance covering both SUA and SIA. Since the net power drawn from the grid $P^{\text{grid}}(t)$ is determined by the supply-demand balance \eqref{eq:power_balance}, it is not part of the action space. The action vector is consequently defined as
\begin{multline}
  a_{t} = \{P^{\text{HVAC}}(t), P^{\text{ESS,c}}(t), P^{\text{ESS,d}}(t), P^{\text{EV,c}}(t), \\
  P^{\text{EV,d}}(t), z_{1}(t), \dots, z_{N}(t)\}.
\end{multline}

\subsubsection{Reward}
Since the primary objective of the agent is to minimise the total energy cost while maintaining indoor thermal comfort and fulfilling EV charging requirements, the instantaneous reward is defined as the negative weighted sum of the associated costs:
\begin{equation}
  \label{eq:reward}
  R_{t} = -\big( w_{g} C_{\text{grid}}(t) + w_{c} C_{\text{comfort}}(t) + w_{d} C_{\text{deg}}(t) \big),
\end{equation}
where the non-negative coefficients $w_{g}$, $w_{c}$, and $w_{d}$ are weights for the relative importance of grid energy cost, comfort violation cost, and battery degradation cost, respectively.

\subsection{Lagrangian SAC-based Energy Management Algorithm}

To solve the constrained energy management problem \eqref{eq:problem_optimisation}, we employ a Lagrangian SAC approach \cite{haarnoja2018soft}. This method transforms the constrained optimisation problem into an unconstrained dual problem via the Lagrangian relaxation method. The objective is to maximise the entropy-regularised expected return while ensuring the expected cumulative cost remains within the safety budget. The Lagrangian objective function is formulated as:
\begin{equation}
  \label{eq:lagrangian_objective}
  \begin{aligned}
    L(\varpi, \lambda)
    &= \mathbb{E}_{\tau \sim \varpi, \xi} \Bigg[ \sum_{t=0}^{\infty} \gamma^t
      \Big( R(s_t, a_t) - \lambda g(s_t, a_t) \\
        &\qquad\qquad\qquad\quad
    + \alpha \mathcal{H}(\varpi(\cdot \mid s_t)) \Big) \Bigg] + \lambda d.
  \end{aligned}
\end{equation}
where $\tau = (s_0, a_0, s_1, \dots)$ denotes the state-action trajectory induced by the policy $\varpi$, and $\mathbb{E}_{\tau \sim \varpi, \xi}$ represents the expectation over these trajectories. The terms $R(s_t, a_t)$ and $g(s_t, a_t)$ correspond to the instantaneous reward and constraint cost, respectively. In this work, we define the constraint cost as $g(s_t, a_t) \triangleq C_{\text{comfort}}(t)$ to penalise comfort violations. The term $\lambda d$ arises from the relaxation of the safety constraint $\mathbb{E}[\sum \gamma^t g(\cdot)] \le d$, where $d$ is the predefined cost limit and $\lambda \ge 0$ serves as the Lagrangian multiplier penalising violations. Additionally, $\alpha$ is the entropy coefficient scaling the entropy term $\mathcal{H}$ to encourage exploration.

We adopt an actor-critic architecture with clipped double Q-learning to mitigate value overestimation. The framework maintains two reward critics $Q_{\phi_1}^{R}, Q_{\phi_2}^{R}$ and two cost critics $Q_{\psi_1}^{C}, Q_{\psi_2}^{C}$. The reward critics approximate the soft Q-value by minimising the Bellman residual:
\begin{align}
  \mathcal{L}(\phi)
  &= \mathbb{E}_{\mathcal{D}} \big[ \big( Q_{\phi}^R(s, a) - y^R \big)^2 \big], \\
  y^R
  &= R(s, a) + \gamma (1-\ell) \Big(
    \min_{j=1,2} Q_{\phi_j'}^R(s', a') \nonumber \\
    &\qquad\qquad\quad
  - \alpha \log \varpi_\theta(a'|s') \Big),
\end{align}
where $\mathcal{D}$ is the replay buffer, $\ell$ is the termination flag, and $\phi'$ denotes parameters of the target network. The cost critic parameters $\psi$ are updated by minimising the MSE loss against a standard Bellman target $y^C$:
\begin{align}
  \mathcal{L}(\psi) &= \mathbb{E}_{\mathcal{D}} \left[ \left( Q_{\psi}^C(s, a) - y^C \right)^2 \right], \\
  y^{C} &= g(s, a) + \gamma (1-\ell) \min_{j=1,2} Q_{\psi_j'}^C(s', a').
\end{align}

The policy $\varpi_\theta$ is updated to maximise the Lagrangian objective in \eqref{eq:lagrangian_objective} via the parametrisation trick. By sampling
$a_\theta(s, \xi) = \tanh(\mu_\theta(s) + \sigma_\theta(s) \odot \xi)$ with $\xi \sim \mathcal{N}(0, I)$, the actor loss becomes
\begin{align}
  \mathcal{L}(\theta)
  &= \mathbb{E}_{s \sim \mathcal{D},\, \xi} \Big[
    \alpha \log \varpi_\theta(a_\theta(s, \xi)\mid s) \nonumber \\
    &\qquad
    - \min_{j=1,2} Q_{\phi_j}^R(s, \tilde{a})
  + \lambda \min_{j=1,2} Q_{\psi_j}^C(s, \tilde{a}) \Big],
\end{align}
where $\tilde{a} = a_\theta(s, \xi)$. This encourages maximising reward and entropy while penalising the weighted constraint cost.

The Lagrange multiplier $\lambda$ is treated as a learnable parameter, updated by dual gradient ascent to enforce the safety budget $d$:
\begin{equation}
  \lambda \leftarrow \big[
    \lambda + \eta_\lambda \big(
      \mathbb{E}_{s \sim \mathcal{D},\, a \sim \varpi_\theta}
      [ \min_{j=1,2} Q_{\psi_j}^C(s, a) ] - d
    \big)
  \big]^+,
\end{equation}
where $\eta_\lambda$ is the learning rate and $[\cdot]^+$ denotes projection onto the non-negative orthant. The entropy coefficient $\alpha$ is tuned online to track a fixed target entropy, avoiding premature convergence to suboptimal deterministic policies.

\section{Simulation Results}

In this section, we evaluate the proposed energy scheduling policy using a simulation of a Swedish household with practical data. To assess its performance, the proposed algorithm is benchmarked against two rule-based policies.

\subsection{Simulation Setup}

\begin{table*}[tb]
  \centering
  \caption{Parameters of household appliances.}\label{tab:appliances_data}
  \begin{tabular}{cllrrrrrr}
    \hline
    \hline
    Index $i$ & Load & Type & Time $t_i^b$ & Time $t_i^e$ & Duration & $L_i^{\text{on}}$ & $L_i^{\text{off}}$ & Power [W] \\
    \hline
    1 & Dishwasher & SUA & 19:00 & 22:00 & 40 mins & - & - & 2,000 \\
    2 & Washing Machine & SUA & 18:00 & 23:00 & 2 hours & - & - & 1,500 \\
    3 & TV & SUA & 19:00 & 23:00 & 1.5 hours & - & - & 200 \\
    4 & Electric Oven & SUA & 18:00 & 20:00 & 0.5 hour & - & - & 3,500 \\
    5 & Robot Cleaner & SIA & 0:00 & 24:00 & 2 hours & 1 hour & 0.5 hour & 50 \\
    6 & Air Purifier & SIA & 0:00 & 24:00 & 12 hours & 1 hour & 1 hour & 40 \\
    7 & Fridge & UA & 0:00 & 24:00 & 24 hours & - & - & 450 \\
    8 & Lights & UA & 18:00 & 23:00 & 5 hours & - & - & 150 \\
    \hline
    \hline
  \end{tabular}
\end{table*}

\subsubsection{Data and System} The simulation setup utilises realistic data categorised into market, environmental, and device parameters. Regarding electricity pricing, the cost model includes hourly spot prices from the Sweden SE3 bidding zone, grid fees based on Göteborg Energi, and applicable taxes, with the electricity selling price set to $r_s = 0.8 r_b$. For renewable generation and weather, the PV array is rated at a peak power of 6.6 kWp. Time-series data for both PV power and ambient temperature are obtained from the Photovoltaic Geographical Information System (PVGIS) \cite{huld2012new}. In terms of household devices, the HVAC system provides both heating and cooling with a maximum power consumption of $P^{\text{HVAC}}_{\max} = 3000$ W. Its thermal dynamics are governed by parameters $\varepsilon = 0.7$ and $\alpha_{\text{T}} = 125/7$ \cite{constantopoulos1991estia,chen2017butler}, aiming to maintain the indoor temperature within the comfort zone of $20^{\circ}$C--$24^{\circ}$C. Finally, the detailed operational requirements for ordinary appliances are summarised in Table~\ref{tab:appliances_data}. The detailed parameters for the battery degradation models, anode potential fitting, and the stochastic probability distributions for EV user behaviour are summarised in Appendix Tables~\ref{tab:param_degra}--\ref{tab:distance_parameters}.

\subsubsection{Batteries}

We employ an LFP battery for the ESS and an NMC battery for the EV, with detailed ageing parameters provided in Table~\ref{tab:param_degra}. Degradation weights are set to $w_{\text{deg}}^{\text{ESS}} = 28000$ and $w_{\text{deg}}^{\text{EV}} = 36750$, reflecting the higher replacement cost and operational criticality of the EV battery.

The ESS has a nominal energy of 13.5 kWh and an assumed nominal capacity of 200 Ah. The EV features a nominal energy of 70 kWh and an approximate nominal capacity of 175 Ah, inferred from its energy rating and a 400 V system voltage. The maximum charging and discharging powers of the EV battery are both 11 kW, while the ESS provides 8 kW charging and 11.5 kW discharging.

\subsubsection{Efficiencies and SoC bounds}

The charging and discharging efficiencies for both ESS and EV are $\eta^{k,c} = \eta^{k,d} = 95\%$, where $k\in \{\text{ESS}, \text{EV}\}$. To avoid high-SoC dwell and deep discharges, thereby extending battery life, the EV SoC bounds are set to $\mathrm{SoC}^{\mathrm{EV}}_{\min} = 20\%$ and $\mathrm{SoC}^{\mathrm{EV}}_{\max} = 90\%$, and the ESS limits are $\mathrm{SoC}^{\mathrm{ESS}}_{\min} = 10\%$ and $\mathrm{SoC}^{\mathrm{ESS}}_{\max} = 100\%$. The energy consumption of the EV is set to $\eta^{\text{EV}} = 0.18$ kWh/km, representing a medium-sized electric vehicle. The target SoC upon EV departure is set to $\mathrm{SoC}^{\mathrm{EV}}_{\mathrm{req}} = 80\%$.

We use the collected data in 2022 for training and the data in 2023 for testing. The initial states are set to $\text{SoC}^{\text{ESS}}(0) = 80\%$ and $\text{SoC}^{\text{EV}}(0) = 90\%$. Simulations were performed in a MacBook Pro laptop with an M3 Pro chip.

\subsection{Training Results}

\begin{figure}[tb]
  \centering
  \includegraphics[width=0.8\linewidth]{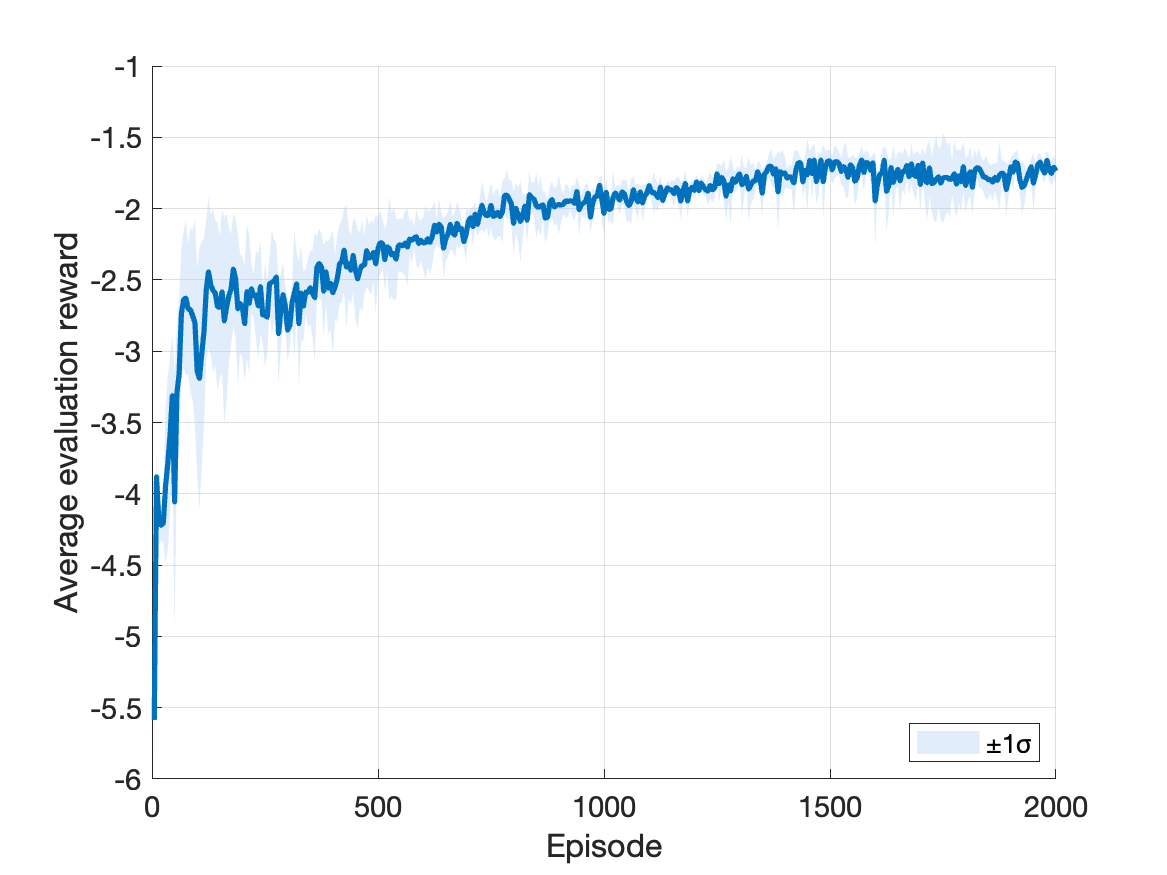}
  \caption{Learning curve of the proposed RL agent.}
  \label{fig:reward_training}
\end{figure}

We train the proposed constrained policy for the HEMS using the developed Lagrangian SAC agent, as outlined in Section~\ref{sec:RL_algorithm}. The training environment follows the Gymnasium API \cite{towers2024gymnasium}, and both the policy and value networks are implemented in PyTorch. Each training episode spans seven consecutive calendar days, with the battery SoC carried over between days to preserve temporal continuity. Every five episodes, we conduct a deterministic evaluation to monitor learning progress. The resulting reward curve is shown in Fig.~\ref{fig:reward_training}. The solid line denotes the mean evaluation reward over five independent training runs, while the shaded region indicates one standard deviation around the mean, capturing the variability due to different random seeds.

\subsection{Energy Optimisation Results}

Upon the completion of 2000 training episodes, the learned policy is evaluated sequentially over the subsequent year.

\begin{figure}[tb]
  \centering
  \begin{subfigure}[b]{0.95\linewidth}
    \includegraphics[width=\linewidth]{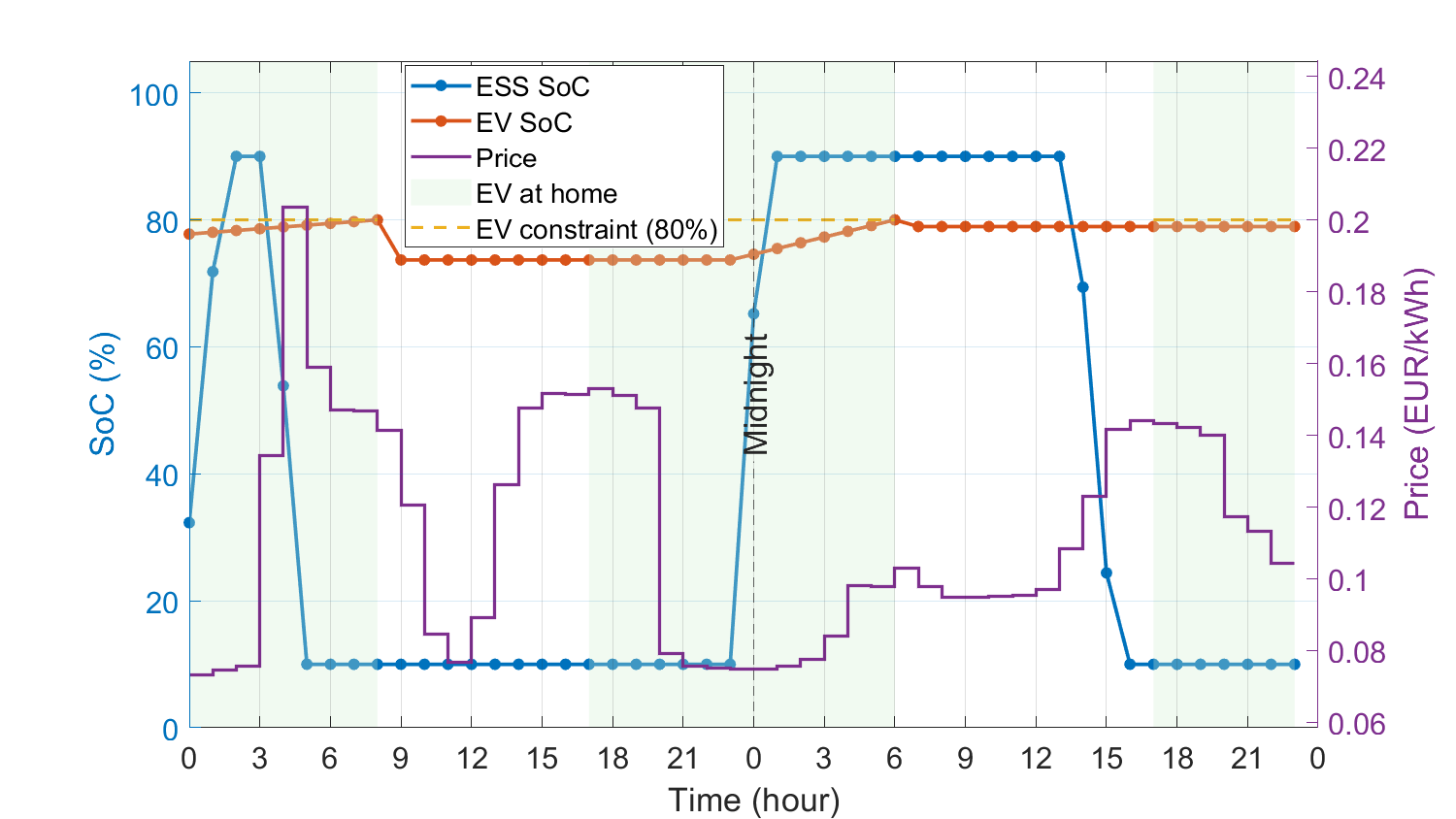}
    \caption{}\label{fig:SOC_proposed}
  \end{subfigure}
  \hfill
  \begin{subfigure}[b]{0.95\linewidth}
    \includegraphics[width=\linewidth]{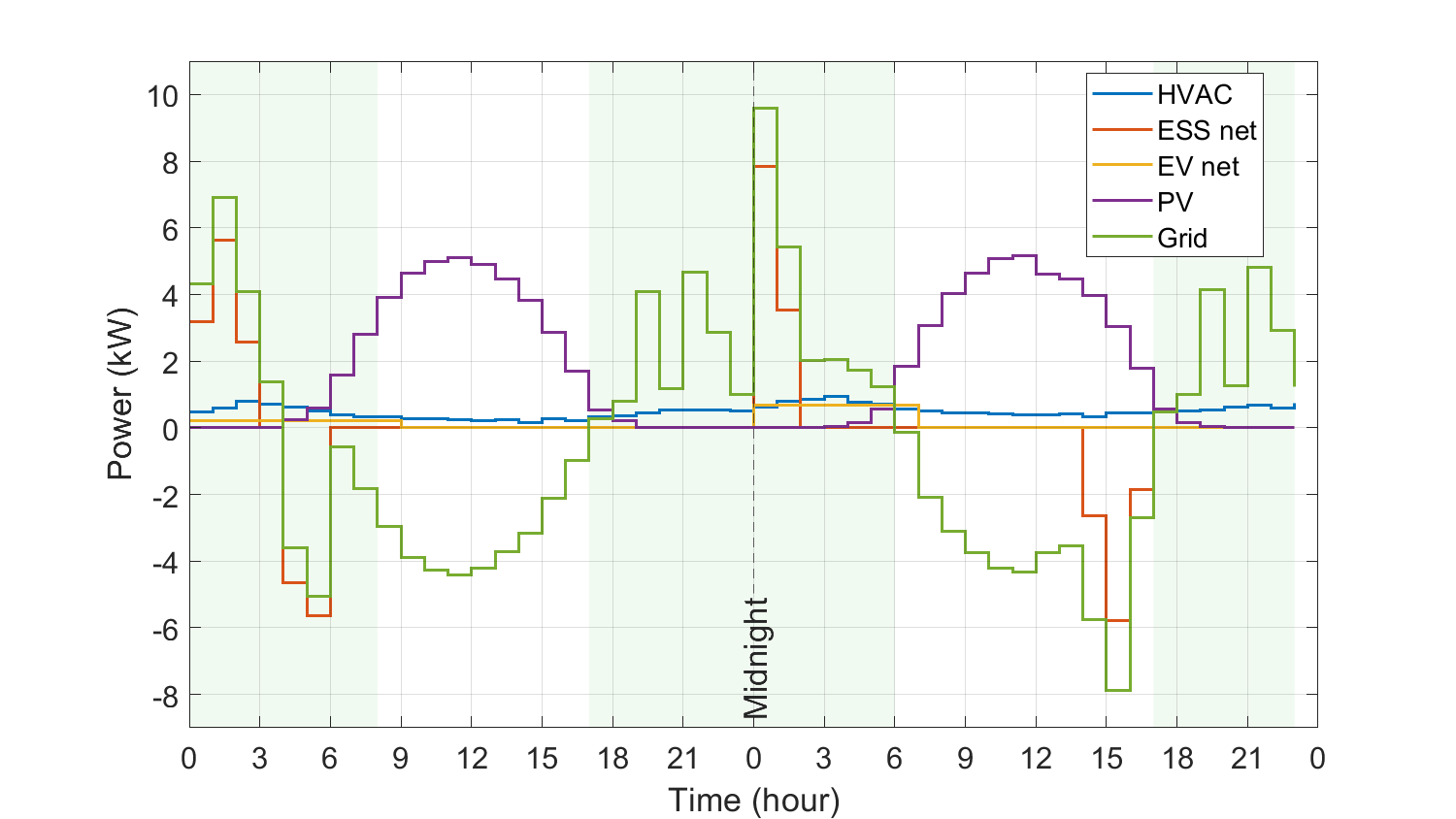}
    \caption{}\label{fig:power_proposed}
  \end{subfigure}
  \hfill
  \begin{subfigure}[b]{0.95\linewidth}
    \includegraphics[width=\linewidth]{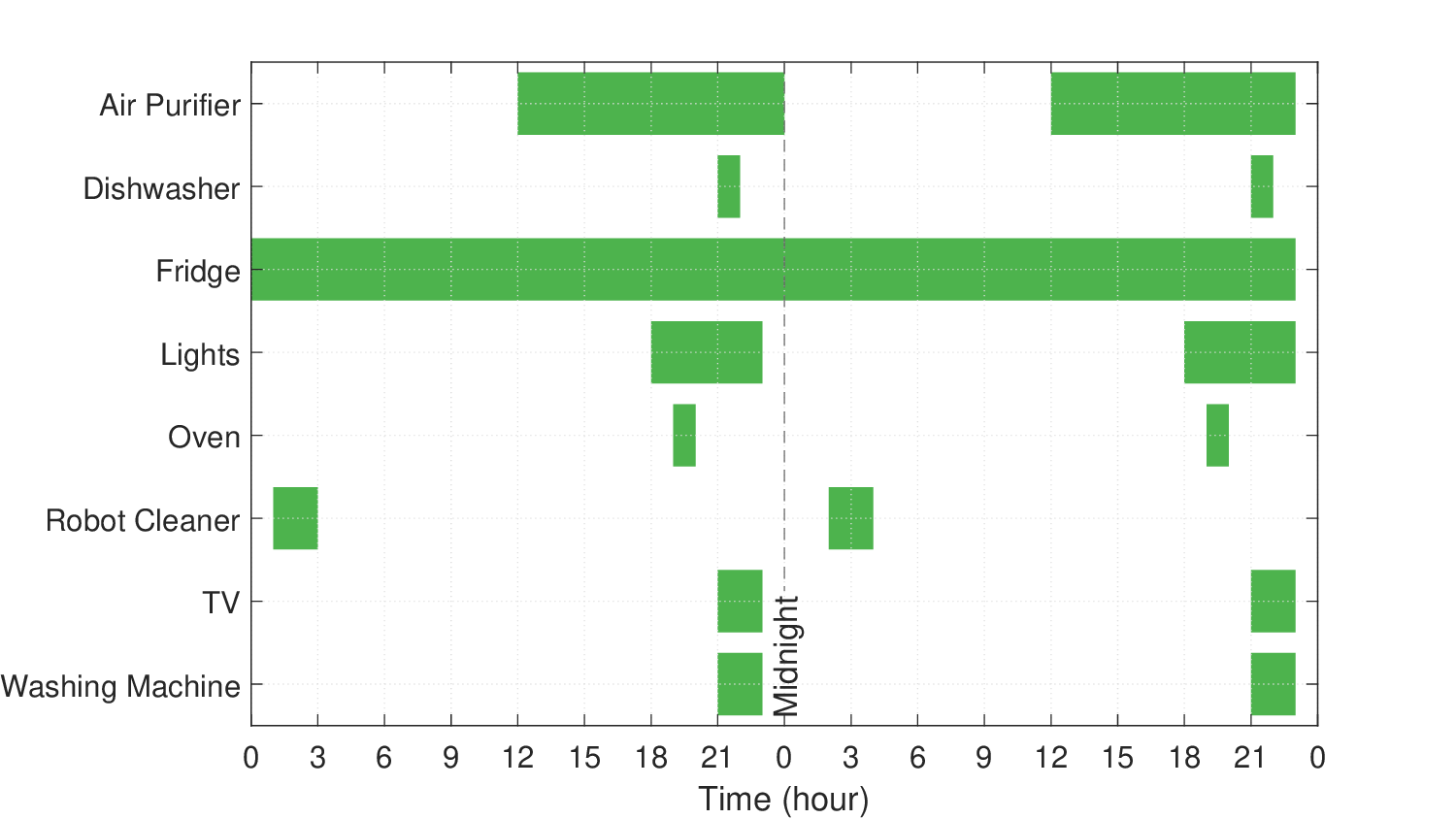}
    \caption{}\label{fig:appliance_proposed}
  \end{subfigure}
  \caption{Energy scheduling performance of the proposed method over a two-day horizon.
    (a) SoC and electricity price,
    (b) Power profiles of the HVAC, ESS, EV, PV, and grid,
  (c) Operating schedules of household appliances. The green shaded area indicates EV at home.}\label{fig:results_proposed}
\end{figure}

The energy scheduling results for a representative day in summer 2023 are illustrated in Fig.~\ref{fig:results_proposed}. Specifically, Fig.~\ref{fig:SOC_proposed} depicts the SoC trajectories of the ESS and the EV alongside the electricity price. The ESS performs price arbitrage by charging during low-price periods, such as early morning, and discharging during peak hours. Meanwhile, the EV successfully reaches the required 80\% SoC upon departure. The power exchange among system components is shown in Fig.~\ref{fig:power_proposed}, where the proposed strategy optimises the utilisation of PV generation and storage to minimise the total operational cost by effectively responding to electricity price variations. Fig.~\ref{fig:appliance_proposed} illustrates the appliance schedules, demonstrating that shiftable loads, including the washing machine, dishwasher, and robot cleaner, are shifted to off-peak periods to minimise operational costs. %

To demonstrate the effectiveness of the proposed framework, we compare its temperature regulation performance and cumulative operating cost against two rule-based baselines. The rule-based strategy employs a reactive HVAC control logic based on outdoor conditions and instantaneous indoor temperature thresholds, utilising dead-bands and hysteresis to mitigate rapid switching. Regarding other components, the ESS performs arbitrage based on fixed price thresholds, the EV charges immediately to the required 80\% without discharging, and appliances operate at the earliest allowable time. The primary distinction between the baselines is that Rule-based 1 permits electricity sales to the grid, whereas Rule-based 2 disables this function.

\begin{figure}[tb]
  \centering
  \includegraphics[width=0.9\linewidth]{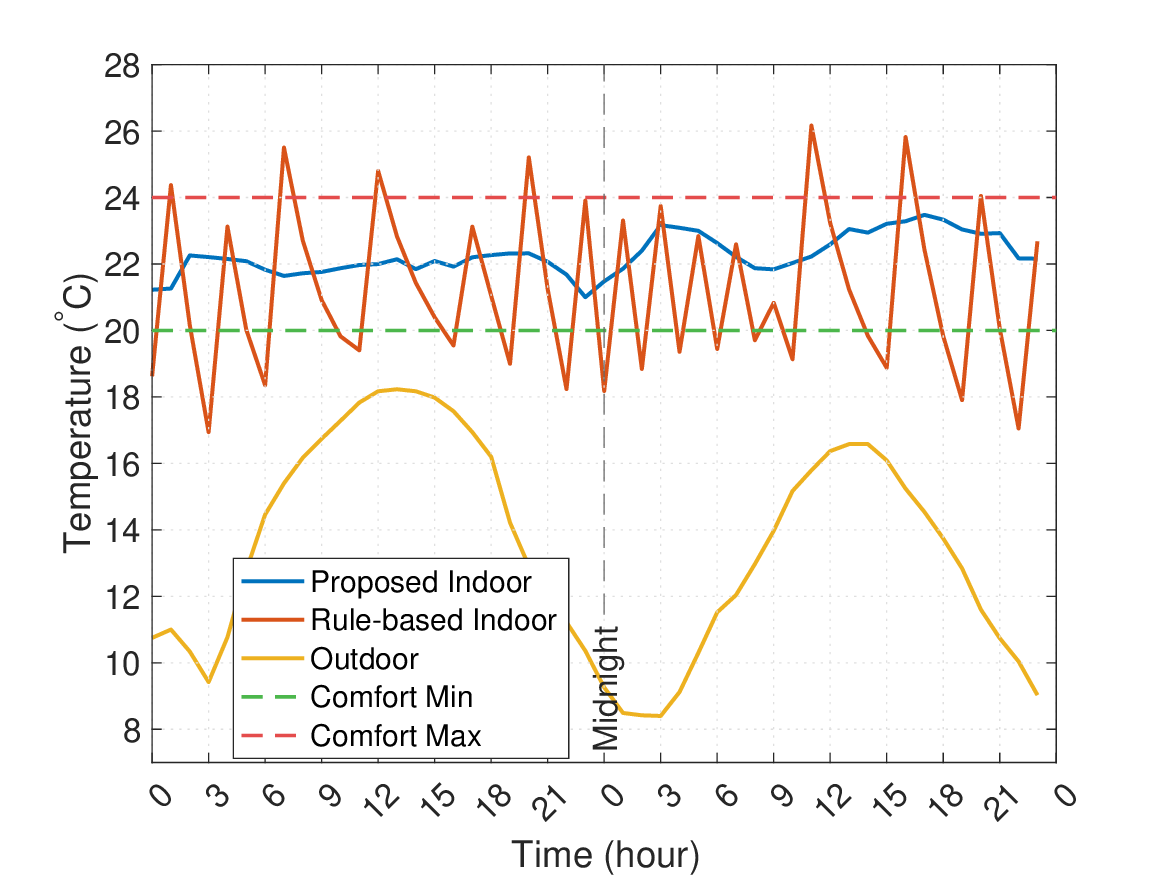}
  \caption{Temperature regulation result based on the proposed method and rule-based approaches.}
  \label{fig:temp_compare}
\end{figure}

As shown in Fig.~\ref{fig:temp_compare}, the proposed framework exhibits superior thermal regulation. The indoor temperature is maintained strictly within the designated comfort band (20$^{\circ}$C--24$^{\circ}$C) with a smooth profile, effectively adapting to external variations. In contrast, the rule-based strategy suffers from high-frequency oscillations and frequent boundary violations, indicating that fixed logic is insufficient for maintaining stability against dynamic disturbances. In addition to improving occupant comfort, the smooth thermal control based on the proposed approach results in a stable power demand for the HVAC system. By alleviating the power surges associated with frequent on-off cycling, this reduces high-rate discharge stress on the battery, which contributes to the lifespan extension as can be seen in Table~\ref{tab:results_comparison}.

In terms of economic efficiency, the proposed method achieves the lowest cumulative operating cost of 1570.78 EUR. As detailed in Table~\ref{tab:results_comparison}, this constitutes a cost reduction of 11.08\% compared to Rule-based 1 and 31.59\% relative to Rule-based 2. The improvement is driven by the optimised coordination of the ESS and EV, which is reflected in the reduced grid and degradation costs. Specifically, the proposed approach improves grid costs by 11.39\% and 33.65\% against the respective baselines, while degradation costs decrease by 8.44\%. These results demonstrate that the proposed algorithm not only guarantees better thermal comfort but also significantly reduces economic expenditure compared to standard heuristic approaches.

\begin{table}[tbp]
  \centering
  \caption{Comparison of annual costs and energy metrics between the proposed method and rule-based approaches}
  \label{tab:results_comparison}
  \resizebox{\linewidth}{!}{
    \begin{tabular}{lrrrrr}
      \toprule
      & Proposed & \multicolumn{2}{c}{Rule-based 1} & \multicolumn{2}{c}{Rule-based 2} \\
      \cmidrule(lr){3-4} \cmidrule(lr){5-6}
      Metric & method [\texteuro] & Value [\texteuro] & Imp. & Value [\texteuro] & Imp. \\
      \midrule
      Grid cost  & 1399.04 & 1578.91 & 11.39\% & 2108.69 & 33.65\% \\
      Degradation cost  & 171.74 & 187.57 & 8.44\% & 187.57 & 8.44\% \\
      Total cost & 1570.78 & 1766.48 & 11.08\% & 2296.25 & 31.59\% \\
      \bottomrule
    \end{tabular}
  }
\end{table}

\section{Conclusion}

In this work, we proposed a RL-based home energy management framework that jointly optimises energy costs, occupant comfort, and energy storage device longevity under uncertainty. A key feature of this study was the explicit integration of heterogeneous degradation characteristics for the stationary battery and EV into the optimisation framework, alongside the inclusion of stochastic EV users behaviour.

Through extensive simulations using real-world data, the proposed strategy demonstrated a strong capability to perform price arbitrage. The net energy purchased from the grid was significantly minimised compared to rule-based baselines, as the system maximised the self-consumption of PV generation and utilised stored energy during high-cost intervals. Quantitative results confirmed the superiority of the proposed approach in reducing the total cumulative operating cost by 11.08\% and 31.59\% compared to two standard heuristic benchmarks, respectively. Furthermore, the approach successfully balanced competing objectives, achieving an 8.44\% reduction in battery degradation costs while maintaining indoor temperature strictly within the comfort bounds, thereby eliminating the oscillations observed in conventional control strategies.

\appendix

This appendix provides the detailed parameter sets utilised in the system modelling and simulation.

\vspace{-1cm}

\begin{table}[tb]
  \centering
  \caption{Parameters for LFP and NCM degradation model.}
  \label{tab:param_degra}
  \begin{tabular}{lcc}
    \hline
    \hline
    Parameter & LFP & NCM \\
    \midrule
    $k_{\text{cal}}$ & $1.9234\times10^{-3}$ & $4.0149\times10^{-4}$ \\
    $E_a$            & $3.0233\times10^{4}$  & $5.9178\times10^{4}$ \\
    $\alpha_{\text{cal}}$         & $-0.05590$            & $-1$ \\[2pt]
    $k_{\text{cyc}}$ & $2.93583\times10^{-6}$ & $4.3131332\times10^{-6}$ \\
    $\alpha_{c1}$    & $1.47611\times10^{-1}$ & $0.3549361$ \\
    $\alpha_{c2}$    & $7.4008\times10^{-3}$  & $1.2308964\times10^{-4}$ \\
    $\alpha_{c3}$    & $0.082035$            & $0$ \\
    $\alpha_{c4}$    & $0.0313111$           & $1$ \\
    $\alpha_{c5}$    & $0.33344256$          & $0.6149392$ \\
    $\alpha_{c6}$    & $331.652158$          & $63.619859$ \\
    \hline
    \hline
  \end{tabular}
\end{table}

\begin{table}[tb]
  \centering
  \caption{Fitted parameters for the anode open-circuit potential model.}
  \label{tab:ua_params}
  \begin{tabular}{c c c c}
    \hline
    \hline
    $i$ & $a_i$ & $b_i$ & $c_i$ \\
    \hline
    1 & $-0.0440$ & $0.1958$  & $0.1088$ \\
    2 & $-0.1978$ & $1.0571$  & $0.0854$ \\
    3 & $-0.6875$ & $-0.0117$ & $0.0529$ \\
    4 & $-0.0175$ & $0.5692$  & $0.0875$ \\
    \hline
    \hline
  \end{tabular}
\end{table}

\begin{table}[t]
  \caption{The parameters of leaving and arriving home distribution}
  \label{tab:time_parameters}
  \begin{centering}
    \begin{tabular}{cc}
      \hline
      \hline
      Parameters & Values\tabularnewline
      \hline
      $\mu_{d}$ & -13.75\tabularnewline
      $\sigma_{d}$ & 0.048\tabularnewline
      $\alpha_{d}$ & 21.05\tabularnewline
      $s_{a}$ & 16.91\tabularnewline
      $\gamma_{a}$ & 0.77\tabularnewline
      \hline
      \hline
    \end{tabular}
    \par
  \end{centering}
\end{table}

\begin{table}[t]
  \centering
  \caption{The parameters of daily distance distribution when $n=3$.}
  \label{tab:distance_parameters}
  \begin{tabular}{cccc}
    \hline
    \hline
    \text{Component $i$} & $A_{i}$ & $\mu_{i}$ & $\sigma_{i}$ \\ \hline
    1 & 0.28 & 2.49 & 1.17 \\
    2 & 0.41 & 7.84 & 4.16 \\
    3 & 0.31 & 26.47 & 25.81 \\ \hline \hline
  \end{tabular}
\end{table}


%

\newpage

\bibliographystyle{IEEEtran.bst}
\bibliography{energy_ref.bib}

@article{yu2019deep,
  title={Deep reinforcement learning for smart home energy management},
  author={Yu, Liang and Xie, Weiwei and Xie, Di and Zou, Yulong and Zhang, Dengyin and Sun, Zhixin and Zhang, Linghua and Zhang, Yue and Jiang, Tao},
  journal={IEEE Internet of Things Journal},
  volume={7},
  number={4},
  pages={2751--2762},
  year={2019},
  publisher={IEEE}
}

@article{xu2023electric,
  title={Electric vehicle batteries alone could satisfy short-term grid storage demand by as early as 2030},
  author={Xu, Chengjian and Behrens, Paul and Gasper, Paul and Smith, Kandler and Hu, Mingming and Tukker, Arnold and Steubing, Bernhard},
  journal={Nature Communications},
  volume={14},
  number={1},
  pages={119},
  year={2023},
  publisher={Nature Publishing Group UK London}
}

@article{safari2011modeling,
  title={Modeling of a commercial graphite/{LiFePO4} cell},
  author={Safari, M and Delacourt, C},
  journal={Journal of The Electrochemical Society},
  volume={158},
  number={5},
  pages={A562},
  year={2011},
  publisher={IOP Publishing}
}

@article{schimpe2018comprehensive,
  title={Comprehensive modeling of temperature-dependent degradation mechanisms in lithium iron phosphate batteries},
  author={Schimpe, Michael and von Kuepach, Markus Edler and Naumann, Maik and Hesse, Holger C and Smith, Kandler and Jossen, Andreas},
  journal={Journal of The Electrochemical Society},
  volume={165},
  number={2},
  pages={A181},
  year={2018},
  publisher={IOP Publishing}
}

@article{lu2020optimal,
  title={Optimal household energy management based on smart residential energy hub considering uncertain behaviors},
  author={Lu, Qing and L{\"u}, Shuaikang and Leng, Yajun and Zhang, Zhixin},
  journal={Energy},
  volume={195},
  pages={117052},
  year={2020},
  publisher={Elsevier}
}

@article{constantopoulos1991estia,
  title={ESTIA: A real-time consumer control scheme for space conditioning usage under spot electricity pricing},
  author={Constantopoulos, Panos and Schweppe, Fred C and Larson, Richard C},
  journal={Computers \& Operations Research},
  volume={18},
  number={8},
  pages={751--765},
  year={1991},
  publisher={Elsevier}
}

@article{chen2017butler,
  title={Butler, not servant: A human-centric smart home energy management system},
  author={Chen, Siyun and Liu, Ting and Gao, Feng and Ji, Jianting and Xu, Zhanbo and Qian, Buyue and Wu, Hongyu and Guan, Xiaohong},
  journal={IEEE Communications Magazine},
  volume={55},
  number={2},
  pages={27--33},
  year={2017},
  publisher={IEEE}
}

@misc{jonkopings_kommun_2024,
author = {{Jönköpings kommun}},
title = {Resvaneundersökning 2024},
year = {2024},
url = {https://www.jonkoping.se/trafik--stadsplanering/resa-och-kollektivtrafik/resvaneundersokning-2024},
note = {Accessed: 2025-08-18}
}

@techreport{regionstockholm2020,
  author      = {{Trafikförvaltningen, Region Stockholm}},
  title       = {Resvaneundersökning 2019},
  institution = {Region Stockholm},
  year        = {2020},
  month       = {Aug},
  url         = {https://www.regionstockholm.se/4a272f/contentassets/d6c4da12e11843c0ab8249c297dfd8fe/resvaneundersokning-2019.pdf},
  note        = {Accessed: 2025-08-18}
}

@article{huld2012new,
  title={A new solar radiation database for estimating {PV} performance in {Europe} and {Africa}},
  author={Huld, Thomas and M{\"u}ller, Richard and Gambardella, Attilio},
  journal={Solar Energy},
  volume={86},
  number={6},
  pages={1803--1815},
  year={2012},
  publisher={Elsevier}
}

@article{farahmandfar2025towards,
  title={Towards net-zero energy buildings: Real-time monitoring, data-driven, and machine learning optimization},
  author={Farahmandfar, Ali and Gharehghani, Ayat and Saray, Jabraeil Ahbabi},
  journal={Energy Conversion and Management},
  volume={343},
  pages={120264},
  year={2025},
  publisher={Elsevier}
}

@article{sangswang2020optimal,
  title={Optimal strategies in home energy management system integrating solar power, energy storage, and vehicle-to-grid for grid support and energy efficiency},
  author={Sangswang, Anawach and Konghirun, Mongkol},
  journal={IEEE Transactions on Industry Applications},
  volume={56},
  number={5},
  pages={5716--5728},
  year={2020},
  publisher={IEEE}
}

@article{gholinejad2022hierarchical,
  title={Hierarchical energy management system for home-energy-hubs considering plug-in electric vehicles},
  author={Gholinejad, Hamid Reza and Adabi, Jafar and Marzband, Mousa},
  journal={IEEE Transactions on Industry Applications},
  volume={58},
  number={5},
  pages={5582--5592},
  year={2022},
  publisher={IEEE}
}

@article{nguyen2013joint,
  title={Joint optimization of electric vehicle and home energy scheduling considering user comfort preference},
  author={Nguyen, Duong Tung and Le, Long Bao},
  journal={IEEE Transactions on Smart Grid},
  volume={5},
  number={1},
  pages={188--199},
  year={2013},
  publisher={IEEE}
}

@article{islam2021coordinating,
  title={Coordinating electric vehicles and distributed energy sources constrained by user’s travel commitment},
  author={Islam, Md Rabiul and Lu, Haiyan and Hossain, M Jahangir and Li, Li},
  journal={IEEE Transactions on Industrial Informatics},
  volume={18},
  number={8},
  pages={5307--5317},
  year={2021},
  publisher={IEEE}
}

@article{saber2022transactive,
  title={Transactive energy management of {V2G-capable} electric vehicles in residential buildings: {An MILP} approach},
  author={Saber, Hossein and Ranjbar, Hossein and Fattaheian-Dehkordi, Sajjad and Moeini-Aghtaie, Moein and Ehsan, Mehdi and Shahidehpour, Mohammad},
  journal={IEEE Transactions on Sustainable Energy},
  volume={13},
  number={3},
  pages={1734--1743},
  year={2022},
  publisher={IEEE}
}

@inproceedings{hussain2020heuristic,
  title={A heuristic-based home energy management system for demand response},
  author={Hussain, Hafiz Majid and Nardelli, Pedro HJ},
  booktitle={2020 IEEE Conference on Industrial Cyberphysical Systems (ICPS)},
  volume={1},
  pages={285--290},
  year={2020},
  organization={IEEE}
}

@article{thomas2018optimal,
  title={Optimal operation of an energy management system for a grid-connected smart building considering photovoltaics’ uncertainty and stochastic electric vehicles’ driving schedule},
  author={Thomas, Dimitrios and Deblecker, Olivier and Ioakimidis, Christos S},
  journal={Applied Energy},
  volume={210},
  pages={1188--1206},
  year={2018},
  publisher={Elsevier}
}

@article{farzin2016practical,
  title={A practical scheme to involve degradation cost of lithium-ion batteries in vehicle-to-grid applications},
  author={Farzin, Hossein and Fotuhi-Firuzabad, Mahmud and Moeini-Aghtaie, Moein},
  journal={IEEE Transactions on Sustainable Energy},
  volume={7},
  number={4},
  pages={1730--1738},
  year={2016},
  publisher={IEEE}
}

@article{erdinc2014smart,
  title={Smart household operation considering bi-directional {EV} and {ESS} utilization by real-time pricing-based {DR}},
  author={Erdinc, Ozan and Paterakis, Nikolaos G and Mendes, Tiago DP and Bakirtzis, Anastasios G and Catal{\~a}o, Jo{\~a}o PS},
  journal={IEEE Transactions on Smart Grid},
  volume={6},
  number={3},
  pages={1281--1291},
  year={2014},
  publisher={IEEE}
}

@article{al2023optimal,
  title={Optimal design of {V2G} incentives and {V2G}-capable electric vehicles parking lots considering cost-benefit financial analysis and user participation},
  author={Al-obaidi, Abdullah Azhar and Farag, Hany EZ},
  journal={IEEE Transactions on Sustainable Energy},
  volume={15},
  number={1},
  pages={454--465},
  year={2023},
  publisher={IEEE}
}

@article{huang2022investigation,
  title={Investigation of electric vehicle smart charging characteristics on the power regulation performance in solar powered building communities and battery degradation in Sweden},
  author={Huang, Pei and Tu, Ran and Zhang, Xingxing and Han, Mengjie and Sun, Yongjun and Hussain, Syed Asad and Zhang, Linfeng},
  journal={Journal of Energy Storage},
  volume={56},
  pages={105907},
  year={2022},
  publisher={Elsevier}
}

@article{wu2016stochastic,
  title={Stochastic control of smart home energy management with plug-in electric vehicle battery energy storage and photovoltaic array},
  author={Wu, Xiaohua and Hu, Xiaosong and Moura, Scott and Yin, Xiaofeng and Pickert, Volker},
  journal={Journal of Power Sources},
  volume={333},
  pages={203--212},
  year={2016},
  publisher={Elsevier}
}

@article{mansour2025deep,
  title={Deep reinforcement learning-based plug-in electric vehicle charging/discharging scheduling in a home energy management system},
  author={Mansour, Shaza H and Azzam, Sarah M and Hasanien, Hany M and Tostado-V{\'e}liz, Marcos and Alkuhayli, Abdulaziz and Jurado, Francisco},
  journal={Energy},
  volume={316},
  pages={134420},
  year={2025},
  publisher={Elsevier}
}

@article{lissa2021deep,
  title={Deep reinforcement learning for home energy management system control},
  author={Lissa, Paulo and Deane, Conor and Schukat, Michael and Seri, Federico and Keane, Marcus and Barrett, Enda},
  journal={Energy and AI},
  volume={3},
  pages={100043},
  year={2021},
  publisher={Elsevier}
}

@article{zenginis2022smart,
  title={Smart home’s energy management through a clustering-based reinforcement learning approach},
  author={Zenginis, Ioannis and Vardakas, John and Koltsaklis, Nikolaos E and Verikoukis, Christos},
  journal={IEEE Internet of Things Journal},
  volume={9},
  number={17},
  pages={16363--16371},
  year={2022},
  publisher={IEEE}
}

@article{wang2016predictive,
  title={Predictive scheduling framework for electric vehicles with uncertainties of user behaviors},
  author={Wang, Bin and Wang, Yubo and Nazaripouya, Hamidreza and Qiu, Charlie and Chu, Chi-Cheng and Gadh, Rajit},
  journal={IEEE Internet of Things Journal},
  volume={4},
  number={1},
  pages={52--63},
  year={2016},
  publisher={IEEE}
}

@article{niu2024uncertainty,
  title={Uncertainty analysis of the electric vehicle potential for a household to enhance robustness in decision on the {EV/V2H} technologies},
  author={Niu, Jide and Li, Xiaoyuan and Tian, Zhe and Yang, Hongxing},
  journal={Applied Energy},
  volume={365},
  pages={123294},
  year={2024},
  publisher={Elsevier}
}

@inproceedings{haarnoja2018soft,
  title={Soft actor-critic: Off-policy maximum entropy deep reinforcement learning with a stochastic actor},
  author={Haarnoja, Tuomas and Zhou, Aurick and Abbeel, Pieter and Levine, Sergey},
  booktitle={International Conference on Machine Learning},
  pages={1861--1870},
  year={2018},
  organization={Pmlr}
}

@article{towers2024gymnasium,
  title={Gymnasium: A standard interface for reinforcement learning environments},
  author={Towers, Mark and Kwiatkowski, Ariel and Terry, Jordan and Balis, John U and De Cola, Gianluca and Deleu, Tristan and Goul{\~a}o, Manuel and Kallinteris, Andreas and Krimmel, Markus and KG, Arjun and others},
  journal={arXiv preprint arXiv:2407.17032},
  year={2024}
}

@article{zhu2024predicting,
  title={Predicting electric vehicle energy consumption from field data using machine learning},
  author={Zhu, Qingbo and Huang, Yicun and Lee, Chih Feng and Liu, Peng and Zhang, Jin and Wik, Torsten},
  journal={IEEE Transactions on Transportation Electrification},
  volume={11},
  number={1},
  pages={2120--2132},
  year={2024},
  publisher={IEEE}
}
\vfill

\end{document}